# Lattice stability and high pressure melting mechanism of dense hydrogen up to 1.5 TPa

Hua Y. Geng,[1,2] R. Hoffmann,[2] Q. Wu[1]

[1] *National Key Laboratory of Shock Wave and Detonation Physics, Institute of Fluid Physics, CAEP; P.O.Box 919-102 Mianyang, Sichuan, P. R. China, 621900*

[2] *Department of Chemistry and Chemical Biology, Cornell University, Baker Laboratory, Ithaca, New York 14853, USA*

## Abstract

Lattice stability and metastability, as well as melting, are important features of the physics and chemistry of dense hydrogen. Using *ab initio* molecular dynamics (AIMD), the classical superheating limit and melting line of metallic hydrogen are investigated up to 1.5 TPa. The computations show that the classical superheating degree is about 100 K, and the classical melting curve becomes flat at a level of 350 K when beyond 500 GPa. This information allows us to estimate the well depth and the potential barriers that must be overcome when the crystal melts. Inclusion of nuclear quantum effects (NQE) using path integral molecular dynamics (PIMD) predicts that both superheating limit and melting temperature are lowered to below room temperature, but the latter never reach absolute zero. Detailed analysis indicates that the melting is thermally activated, rather than driven by pure zero-point motion (ZPM). This argument was further supported by extensive PIMD simulations, demonstrating the stability of *Fddd* structure against liquefaction at low temperatures.

# I. INTRODUCTION

Hydrogen, the simplest element, shows complex behavior under compression [1-6]. It has at least four allotropes in the solid state that were already known, and exhibits an anomalous melting temperature ($T_m$) that peaks at about 100 GPa [7-13] and then decreases downwards [12-15]. It was speculated that at higher pressures dense hydrogen in a metallic state might melt driven not by thermal motion of nuclei (as other elements usually are) but rather by pure nuclear quantum effects (NQE), or equivalently, by the zero-point motion (ZPM) of nuclei [16,17]. This conjecture is tantalizing and hints the possibility of a quantum liquid in its ground state as 0 K is approached [18].

Recent numerical simulations predicted that this descent might continue beyond 1 TPa [19]. However, there are two fundamental questions yet to be answered: (*i*) does dense hydrogen really melt at 0 Kelvin? (*ii*) What are the respective role played by the softening of the interaction potential, as well as that played by the NQE in this decline? Namely, does the low-temperature melting originate from the flatness of the potential energy surface [20] or simply because of the enormous ZPM? This query is important, because an analogous decrease of $T_m$ has also been observed in the alkali metals such as Li [21,22] and Na [23], where NQE is insignificant. For these two elements, the $T_m$ rises again at higher pressures. Considering the similarity of metallic hydrogen ($H_M$) with the alkali metals [24], it is reasonable to expect that hydrogen should also follow a similar trend. A consequent supposition is that the potential softening could be limited, and the energy surface (ES) of $H_M$ in this pressure range might still be rough, with noticeable energy wells and barriers. If true, this will provide profound insight into the phase stability of solid $H_M$, because thermally driven forces will diminish with decreasing temperature if the destabilization (or melting) of a crystal is thermally activated (TA). On the other hand, when near the groundstate, the only possible dynamical forces that can destabilize a lattice are ZPM or quantum tunneling, the latter a mechanism in quantum melting that has received some attention only very recently [25]. In this hypothetical scenario, the particle tunneling length and the height and width of the barriers on the ES are the key parameters that dictate the melting behavior.

In this article, we will demonstrate for the first time that within the pressure range from 500 to 1500 GPa, $H_M$ does fall in this regime (*i.e.*, with limited softening in the potential) and have noticeable energy barriers. One of the consequences is a strong meta-stability of crystalline or glass phases at low temperatures. Furthermore, the solid groundstate of dense hydrogen has also been established at the level of density functional theory (DFT) with the first direct numerical evidence obtained by using extensive AI-PIMD simulations.

## II. METHOD AND THEORETICAL DETAILS

**A. First-principles calculations**

In our calculations, the many-body electron problem is treated with DFT, and periodic boundary conditions (PBC) are used to model the solid and/or liquid phases. AIMD simulations are carried out in a micro canonical ensemble (*NVE*), in which the particle number *N*, internal energy *E*, and cell volume *V* are conserved quantities. The classical melting is modeled using the "Z-curve" method [26], in which the internal energy is adjusted by initializing different temperatures in the system. By gradually increasing *E* with the cell volume being fixed, the solid phase evolves into a superheated region, and then abruptly collapses to a liquid state after reaching a critical point. The thermodynamic condition immediately after the structural collapse gives exactly the melting pressure and temperature [26]. The time step for integration of the classical motion equations is 0.5 fs. A typical AIMD simulation runs 6000 time steps, corresponding to 3 ps. Note that the structure change and melting in dense hydrogen usually take place within 1 ps in *classical* MD simulations. The final pressure and temperature are obtained by statistics over the last 2000 time steps.

In AI-PIMD simulations, the quantum motion of protons is taken into account through the path integral formalism of quantum statistical mechanics [27-29]. When evaluating the NQE in the superheating limit, eight beads along the imaginary time line are used to approximate the Trotter decomposition of the propagators. But 32 beads are also used at a pressure of ~1.5 TPa, to check the convergence of the path integral at around 300 K. In some cases, for example *Fddd* at 100 K and ~700 GPa, more beads are used to check the impact of bead number on the (meta-)stability of the solid phases. Two-phase simulation is carried out with 32 beads, whereas the enthalpies at 50 K and ~1.5 TPa are also calculated with 64 beads. The shortest propagator is estimated by the primitive approximation [27]. It corresponds to classical motions at 2400 K for the case with $T$=300 K and 8 beads, and 3200 K for the case with 64 beads at 50 K, which is accurate enough for our current purpose. All AI-PIMD simulations are conducted in the *NVT* ensemble, and the superheating limit is estimated by using the *heat until melting* strategy. The melting temperature is estimated by using the NQE corrected superheating limit, together with the classical superheating degree, and the classical melting temperature difference between various k-point meshes. Alternatively, two-phase method is also used to estimate the melting temperature. Most AI-PIMD simulation runs to 5 ps, with the last 1 ps taken for thermodynamic properties statistics. In two-phase simulations, however, longer simulation time is used, to ensure the structural equilibrium, where the averaged

AI-PIMD simulation time is between 7 to 10 ps. As usual, we did not include exchange operations in AI-PIMD simulations, since protons are well separated from each other even in the liquid phase at 50 K, as [19] reported.

In both AIMD and AI-PIMD we use the same simulation cell, if without specific statement, 480H/cell for *Fddd* and liquid phase, and 432H/cell for Cs-IV phase, respectively. The forces required in the equations of motion for protons in both AIMD and AI-PIMD are calculated by density functional theory, using VASP—a code based on plane-wave methods [30]. The projector augmented-wave (PAW) pseudo-potential is employed to describe the proton-electron interactions [31,32]. The Perdew-Burke-Ernzerhof (PBE) [33] parameterizations for the electron exchange-correlation energy functional are used. It is worth mentioning that previous work revealed that semi-local exchange-correlation functional might not be enough for accurate calculation of properties of molecular phases of dense hydrogen, mainly due to the poor description of the van der Waals interactions. However, after hydrogen dissociates into an atomic phase, comparative studies showed that PBE works well in this regime [13,34-36], which is exactly the region we are interested in.

The potential energy surface is generated with various k-point sampling meshes (KPM). A case of two high symmetry special k-points (2KP) is adopted to sample the Brillouin zone: one is the gamma point and the other at half along the <111> direction of an orthorhombic cell (the high-symmetry point *R*); they are reweighted so as to give a best description of the total energy and pressure, in a spirit analogous to Baldereschi mean value point [37-39]. Besides this, regular meshes with a size varied from $2 \times 2 \times 2$ up to $4 \times 4 \times 4$ are also used. The convergence of the k-points is carefully checked, which shows that the total energy and pressure are fully converged with a $3 \times 3 \times 3$ mesh.

Unless specifically noted, most AIMD simulations are carried out with the $3 \times 3 \times 3$ k-points mesh; whereas due to the computational cost, AI-PIMD are usually done with a mesh of $2 \times 2 \times 2$, and a correction of $A = A_{2\times2\times2} + (A_{3\times3\times3} - A_{2\times2\times2})_{\text{AIMD}}$ is applied when necessary. The cutoff for the kinetic energy of the plane-wave basis is 600 eV, which is high enough for MD simulations. Increasing this energy cutoff to 800 eV does not give different results. This setting of the DFT computational parameters produces a stress tensor (as well as the pressure) with an uncertainty less than 1 GPa, which is good enough for our purpose here.

**B. Projected pair correlation function**

Angularly averaged pair correlation function (PCF) is a powerful tool to detect structural changes. It is defined as

$$g(r) = \frac{V}{N^2} \sum_{i \neq j}^{N} \delta(r_{ij} - r), \tag{1}$$

where $r_{ij}$ denotes the distance between particle $i$ and $j$. In a homogeneous liquid, $g(r)$ becomes the well-known radial distribution function. This function, as the prefix "angularly averaged" implies, removes all orientation dependence and the anisotropy of a solid system. The projected PCF, which we will define now, on the other hand, is an attempt to bring the underlying anisotropy back, while keep the simplicity of the mathematical operations. This function is valuable for us in discovering an exotic new phase of $H_M$ that is anisotropy but flowing like a liquid.

The projected PCF along direction $\vec{k}$ is defined as

$$G_{\vec{k}}(\rho) = \frac{S}{N^2} \sum_{i \neq j}^{N} \delta(r_{ij}^{\perp} - \rho). \tag{2}$$

Here $S$ is the projection area, and the distance between particle $i$ and $j$ on the projection plane is $r_{ij}^{\perp} = \|(\vec{r_i} - \vec{r_j}) \cdot (\mathbf{1} - \vec{k}\vec{k})\|$, where $\|\cdots\|$ denotes taking the vector length. Obviously, if the system is two-dimensional and perpendicular to the projection direction, then $G_{\vec{k}}(\rho)$ is identical to the angularly averaged PCF $g(r)$ on that plane. It is worth noting that projected PCF depends on the geometry of the projected region. However, for the regular orthorhombic cell, projection along the Cartesian directions always gives well-defined results.

Specifically, if the system is a homogeneous liquid, then one can derive a simple relation between $g(r)$ and $G(\rho)$. Considering a reference particle, all other particles surround it with a distribution function given by $g(r)$. This can be viewed as being composed by a series of spherical shells. Then the projected PCF can be obtained by the following identity (derived from particle conservation)

$$\frac{2\pi\rho\Delta\rho}{S} G(\rho) = \frac{4\pi}{V} \int_{r>\rho} r^2 g(r) \frac{\Delta S(\rho)}{4\pi r^2} dr, \tag{3}$$

where $\Delta S(\rho)$ is the area of the infinite thin strips on the spherical shells that are perpendicular to the projection direction and have a radius of $\rho$. Simple geometrical analysis gives $\Delta S(\rho) = 4\pi r\rho\Delta\rho/\sqrt{r^2 - \rho^2}$, thus we have

$$G(\rho) = \frac{2S}{V} \int_{r>\rho} \frac{rg(r)}{\sqrt{r^2 - \rho^2}} dr. \tag{4}$$

Projection of a series of spherical shells onto a plane is not as simple as the projection of an orthorhombic cell: the geometry factor $S/V$ is difficult to determine here. For practical purpose, we cut the shells by using a cylinder with equal height and diameter, and then project the shells within the cylinder onto its base plane. The thickness of the projected region generated in this way is about the same order of the cell dimension as our MD simulations. An instructive example for the application of projected PCF is given in the Supplementary Information (SI) [40]. Here we only note that if the system is anisotropic, then the projected PCFs along different direction will show different behavior; and if there is long-range ordering then the projected PCF will have distinct features. In contrast, the projected PCF of homogeneous liquid is independent of projection direction, with a simple feature of monotonic increasing of $G(r)$ to the first peak and then quickly growing featureless at larger distances.

**C. Richardson extrapolation**

In the primitive approximation of path integral, the dependence of the integrated quantities such as the energy on the number of beads $N_b$, is scaled as [41]

$$E = E_0 + A_2 N_b^{-2} + A_4 N_b^{-4} + \cdots \tag{5}$$

when $N_b \to \infty$. Therefore extrapolation of AI-PIMD results evaluated at finite $N_b$ to the infinite one can be done using the Richardson scheme, which works well for most system when $N_b$ is large enough [41]. The extrapolation formula is

$$E_\infty = E_2 + \frac{(N_{b1}/N_{b2})^2}{1 - (N_{b1}/N_{b2})^2}(E_2 - E_1). \tag{6}$$

We extrapolate the AI-PIMD internal energy and pressure with this formula using 32 and 64 beads, respectively. We believe these values of $N_b$ are large enough, and they are the most accurate calculations that can be done with our currently available computational resources. Comparing the enthalpy difference calculated with 32 and 64 beads at 50 K and 1.5 TPa, we found that they are not qualitatively different. Therefore we are confident in this setting, and believe that this extrapolation provides at least qualitatively correct results, which is enough for our current purpose.

**D. Enthalpy correction**

Most of our calculations are performed at constant volumes, thus the resultant pressure is slightly different for different phases with various number of beads. In order to compare the relative stability and to align the enthalpy at the same pressure, following correction to the enthalpy has been made

$$H(P) = H(P_0) + \Delta P V_0 - \frac{\Delta P^2 V_0}{2B(V_0)}. \tag{7}$$

Here the pressure difference with respect to a given volume $V_0$ is $\Delta P = P - P_0$. The bulk modulus is taken as 3.4 TPa, which is a good estimate for dense hydrogen at the studied pressure range [20]. It should be pointed out that the correction is insensitive to the bulk modulus for our interested cases here, and the third term was found to be insignificant. As can be seen in the Fig.6 that will be shown below, the correction is almost linear, indicating the reliability of this approximation to the enthalpy.

**E. Potential energy surface exploration**

In quantum mechanics, the motion of protons is governed by the Hamiltonian $H = T_I + U(\boldsymbol{R})$ within the Born-Oppenheimer approximation, where $T_I$ is the nuclear kinetic operator and $U(\boldsymbol{R})$ the potential energy surface felt by nuclei. One of our primary purposes in this article is to characterize the features in $U(\boldsymbol{R})$, and its role (as well as that played by $T_I$) on the cold melting of dense hydrogen. In order to approach the thermodynamic limit, we employ a large enough simulation cell (480H for *Fddd* and 432H for Cs-IV phase, respectively). To explore the energy surface of such a big system directly is an insurmountable task. Fortunately, since we care about only the main characteristics of the energy surface, we can take the advantage of the fact that in a classical system with conservative force fields, the probability for an equilibrated system to jump out of a potential well (and to overcome an energy barrier as well) is roughly proportional to the temperature. Using this property one can extract the desired information from classical AIMD simulations.

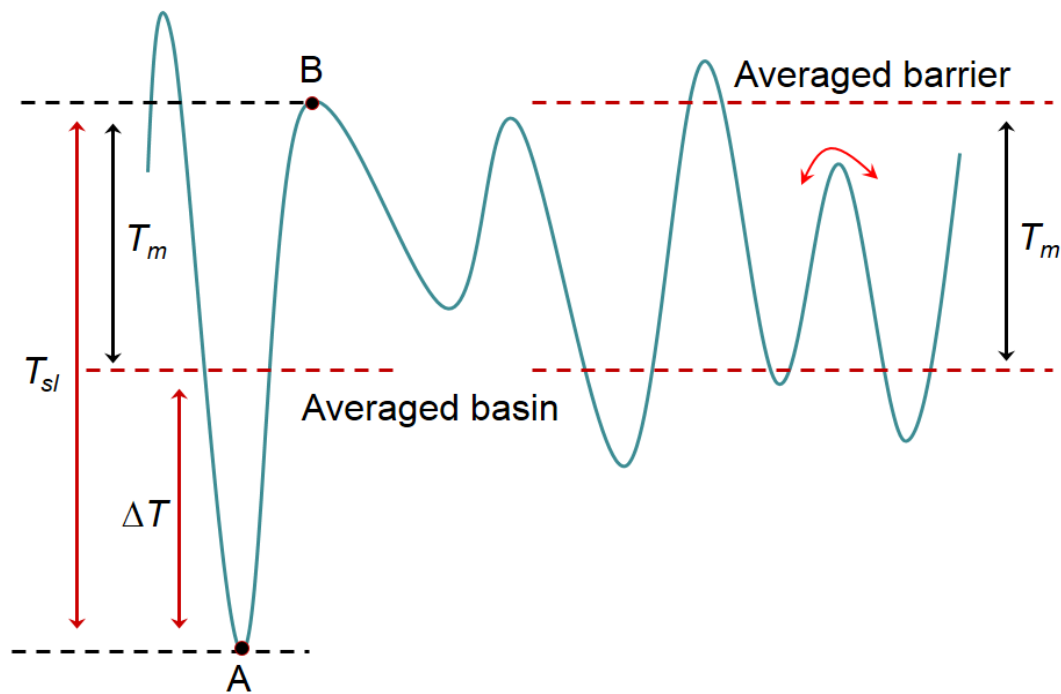

FIG. 1: (color online) Schematic of an energy surface and the strategy of using classical kinetic energy (or equivalently the equilibrium temperature) to explore the surface's main characteristics. Note that classical $T_m$ provides a practical estimate of the averaged barrier height, and $T_{sl}$ gives an assessment of the well depth of the initially solid phase.

Figure 1 illustrates the principle and the strategy we used to assess $U(\boldsymbol{R})$. Starting from a given solid phase (point A), by gradually heating the system, the crystal will fluctuate and finally reach point B where it cannot resist the thermal disturbance anymore, consequently collapsing into other solid or liquid phases. The temperature at this point corresponds to the superheating limit (SL) $T_{sl}$, which can be viewed as an effective measure of the potential well depth. Similarly, immediately after collapsing into a liquid phase, the equilibrium temperature (*i.e.*, the classical $T_m$) gives the minimal classical kinetic energy that is required for the system to travel freely across all underlying barriers in the energy surface. Therefore $T_m$ can be taken as a measure of the averaged height of the energy barriers surrounding the initial solid phase. The difference $\Delta T = T_{sl} - T_m$, or the superheating degree (SD), gives a simple estimate of the static energy difference between the liquid and the initially solid phase.

The "Z-curve" method [26] just mentioned above, in which it is the internal energy rather than the temperature that is tuned, is an ideal tool for this purpose. At the point of lattice collapsing the internal energy of the solid and the liquid should be equal, *i.e.*, $E_s(T_{sl}) = E_l(T_m)$. Destruction of the crystalline structure leads to a redistribution of this energy between kinetic and potential parts, thus changes the equilibrium temperature and pressure accordingly. Though Z-method is simple to use and usually works well, it was also reported that sometimes it overestimates the $T_m$ by up to 30% [42,43]. It is interesting to notice that almost all of these reported cases are related to small simulation cell and heavy elements. In order to examine the performance of Z-method in dense hydrogen, we calculate the melting temperature of *Cmca*-4 phase. The Z-method result is 581 K at 310 GPa, in a perfect agreement with Liu *et al.*'s two-phase method (using *NPT* ensemble) result of 580 K at 300 GPa [15]. Therefore we conclude that it is unlikely that our method used here will have large overestimation of the classical superheating limiting and melting temperature in dense hydrogen.

## III. RESULTS AND DISCUSSIONS

**A. Limited potential softening**

In our AIMD calculations, both Cs-IV [44] and *Fddd* phases are used as the solid candidates. *Fddd* is a low-symmetry distortion of Cs-IV and degenerate in enthalpy with the latter as the (currently proposed) least enthalpy crystalline phase [20] of $H_M$ in the pressure range studied (See SI for their structural connections [40]). Inclusion of *Fddd* has two purposes: (*i*) to improve the reliability of the computational results by coverage of a broad low-lying phase space; (*ii*) because of the geometric connection between these two structures, there might be dynamic oscillations between them, which, if observed, are a precursor of quantum melting [20].

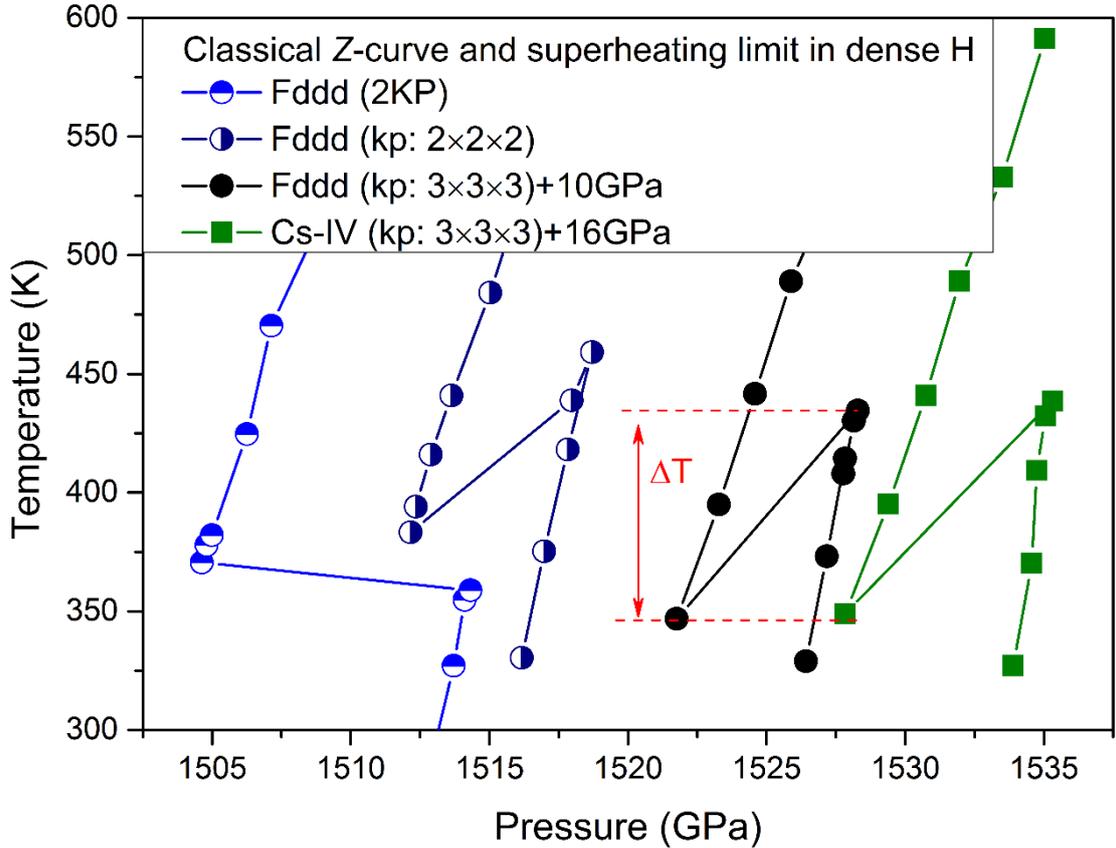

FIG. 2: (color online) Typical Z-curves in the *P-T* plane for $H_M$ at around 1.5 TPa calculated with AIMD simulations using the *NVE* ensemble and different k-point meshes.

Typical Z-curves calculated at ~1.5 TPa are shown in Fig.2. The indication is that the estimated initial well depth is about 450 K (or 39 *m*eV), and the averaged barrier height underlying the liquid phase is ~350 K (or 30 *m*eV) [45]. In terms of the static energy, the initial Cs-IV or *Fddd* phase is favored by about 18 *m*eV (in average) against transient structures in the liquid phase. This is in line with previous static lattice calculations, where an enthalpy difference of the order of ten *m*eV/H among low-lying structures was reported [20,24]. Different resolutions in k-point mesh (KPM) are also examined. As can be seen from Fig.2, this changes the energy surface moderately, especially in the case of two special k-points (2KP) where the relative stability of solid phases has been qualitatively changed (as indicated by the negative $\Delta T$).

It is necessary to point out that we did not observe any phase fluctuations. A single-way transition from Cs-IV to *Fddd* does occur in the case of 2KP, but it is not an oscillation. The same conclusion also holds in AI-PIMD simulations, in which the NQE has been included. This observation implies that the precursor of a quantum melting is difficult to achieve. It also suggests that when approaching the thermodynamic limit, *Fddd* and Cs-IV are distinct phases, and their respective basins in the phase space do not merge into a single one [20].

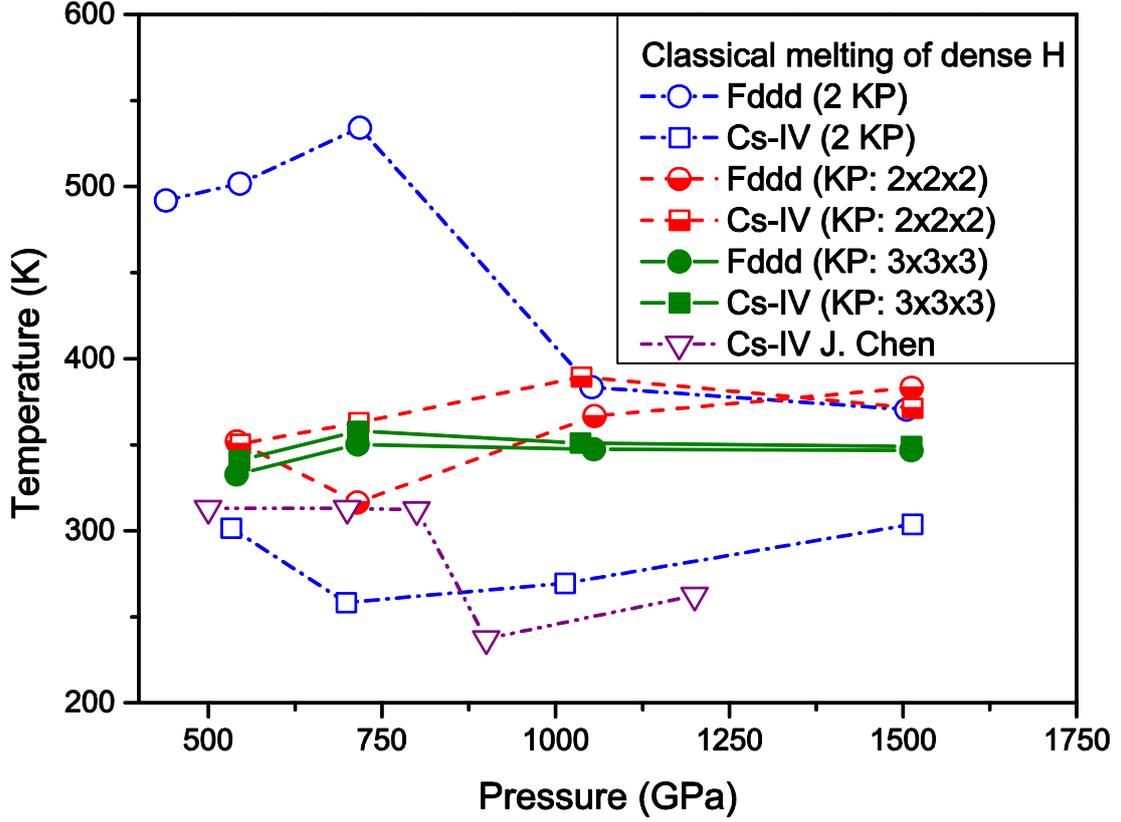

FIG. 3: (color online) Convergence of the classical melting curve of $H_M$ calculated with Z-curve method using AIMD simulations in the *NVE* ensemble. Notice the large error that resulted from insufficient KPM sampling. The data of J. Chen are from [19].

The convergence in our estimated energy surface shape can be inferred from Fig.3, in which the classical $T_m$ as a function of pressure and its variation with respect to different k-point meshes are plotted. It can be seen that the energy surface converges with a 3×3×3 k-points mesh or higher for the cell size we used. The deviation in the $T_m$ of *Fddd* from that of Cs-IV phase, especially for those calculated with low k-points meshes, is a strong indicator that these two phases are physically distinct. By comparison, the results reported by J. Chen *et al.* [19], as our 2KP case here, underestimated the energy barrier by a magnitude of 50~100 K, thus underestimated the stability of the Cs-IV phase as well. When the pressure increased from 500 GPa to 1.5 TPa, our calculation shows that the density of $H_M$ increases 56%, and the averaged inter-atomic distance reduced by ~0.18 Å. This volume shrinkage, however, does not change the main characteristics of the energy surface very much. An important information conveyed by this (about which we did not have any knowledge before) is that the potential softening in $H_M$ is limited, and the main features of energy surface (*e.g.*, the averaged barrier height) have a very weak pressure dependence, which is corroborated by the flatness in the calculated classical $T_m$.

**B. Nuclear quantum effects**

*1. Assessment with perfect lattice*

Above AIMD analysis revealed two important facts: (*i*) though *Fddd* and Cs-IV phases are distorted structures with each other, they respectively have independent basins, and thus are distinctly different phases; (*ii*) the potential softening in $H_M$ is limited, and reaches a flat level when above 500 GPa. With this insightful understanding one can concludes that any further descent in $T_m$ must be because of the nuclear quantum effects. Now we turn to discuss how NQE lowers the melting temperature.

It is well known that in addition to the barriers in the potential energy surface that determine the degree of difficulty for a system to travel from one coordinate configuration into another, destabilization or melting of a lattice is also governed by kinetic operator $T_I$, which generates the dynamical driving forces to overcome the energy barriers. This gives rise to the thermal noise in the classical case, and the NQE in a quantum one. The former depends only on the temperature, whereas the latter is also affected by nuclear masses and localization of the wave function, and manifests itself in ZPM and/or tunneling. From the prospect discussed above, the continuous descent of $T_m$ beyond 500 GPa as predicted in [19] must be a consequence of NQE. There are three mechanisms by which NQE can lower the $T_m$: (1) quantum motion of nuclei leads to a correction term to the free energy of the solid and liquid phases (*e.g.*, zero point enthalpy), thus changes their equality position; (2) the potential well of the solid phases is too shallow to hold the eigenstates of lattice vibrations, resulting in spontaneous delocalization of the nuclear wavefunction; (3) identical particle statistics, *i.e.*, exchanges of identical particles, further contributes to the free energy of the liquid phase, and also enhances the probability for particles to tunnel through the potential barriers [25]. Within our studied temperature range that is above 50 K, exchange in the liquid phase is negligible [19], thus in the following we do not consider case (3), and only the first two mechanisms will be investigated.

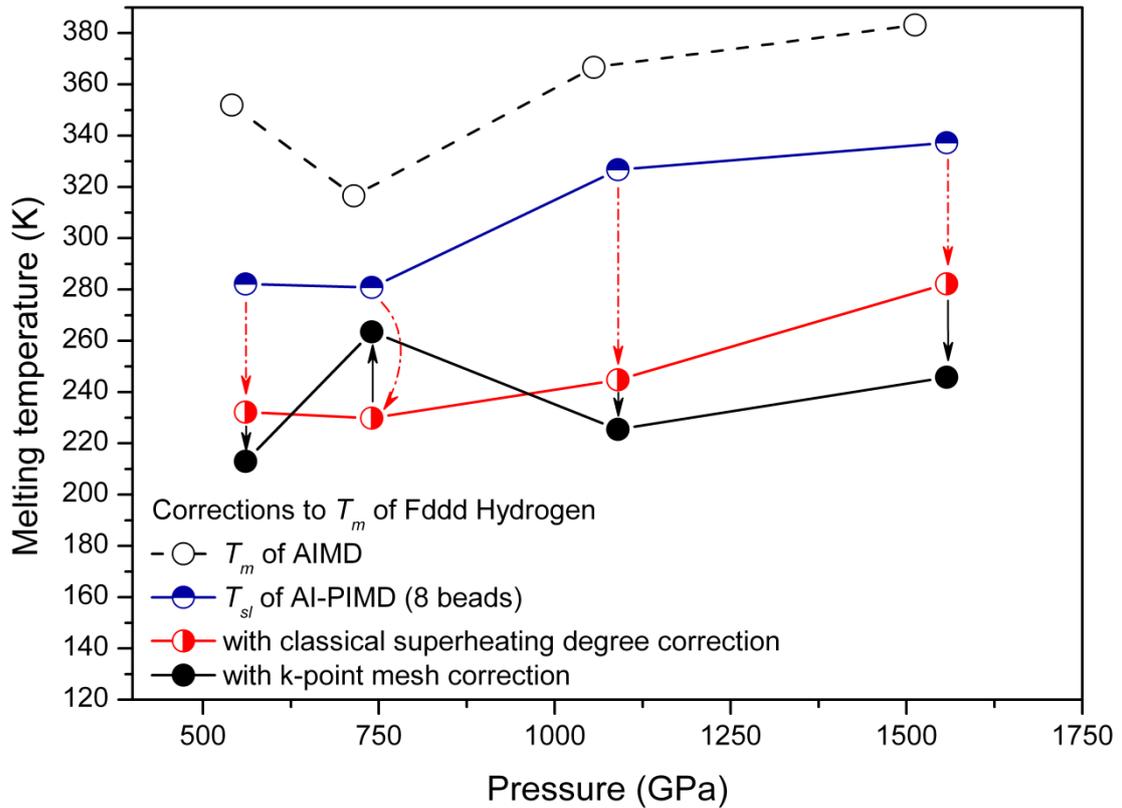

FIG. 4: (color online) Comparison of the melting curves of $H_M$ calculated with AIMD and AI-PIMD simulations. The latter uses eight beads to capture the NQE on superheating limit. Corrections with respect to classical superheating degree and k-point meshes are also plotted.

The NQE on the superheating limit can be estimated by AI-PIMD simulations [28,29]. Since this occurs at relatively high temperatures, only 8 beads were used to discretize the integral path. At 1.5 TPa, $T_{sl}$ obtained in this way is about 330 K. Using 32 beads slightly lowers the $T_{sl}$ to 310 K. This small change indicates that it is adequate to use 8 beads for this purpose. In contrast, the classical $T_{sl}$ is ~450 K. One simple and crude way to assess the NQE corrected $T_m$ from $T_{sl}$ is by subtracting from it the classical superheating degree $\Delta T$. The results obtained are shown in Fig.4 by comparison to the classical $T_m$ calculated with AIMD using the same DFT setting. Also shown is the further correction to account for the convergence of k-points meshes. Note that these results are below room temperature, but higher than 200 K. The high $T_{sl}$ indicates that the NQE has limited effects to *destabilize* the lattice. Or to put it in other words, if the actual $T_m$ is at ultra-low temperatures as reported in [19], the solid phases should have strong meta-stability against melting.

An inference from Fig.4 is that there is no spontaneous delocalization of the nuclear wavefunction, hence the second mechanism mentioned above is disproved. This suggests that melting of $H_M$ in this pressure range is thermally activated, and can be described by adding a quantum correction term to the free energy functional of a classical model. To solidify this argument, one needs additional calculation to show that the solid phases are robust against spontaneous quantum melting, especially at low temperatures where protons have long de Broglie thermal wavelength, thus long tunneling length. For this purpose, we carried out direct AI-PIMD simulations of the *Fddd* phase at 100 K under 700 and 1000 GPa using 24 beads. The results indeed show that *Fddd* is at least metastable under these conditions. Using the same setting as [19] (200H/cell) and increasing the number of beads from 32 to 36, we also checked the Cs-IV phase at 1 TPa and 100 K, which confirms the (meta-)stability of Cs-IV phase against spontaneous melting. Extensive calculations using 128 beads also prove the (meta-)stability of solid $H_M$ at 50 and 100 K under 700 GPa, in which due to the exceptional computation cost only 36 H/cell and 32 H/cell were used for the Cs-IV and *Fddd* phase, respectively. At 1.5 TPa and 50 K, long enough AI-PIMD simulations with 64 beads also confirm that solid *Fddd* phase is stable. In all of these simulations, no tunneling event was observed. This finally establishes the thermally activated melting mechanism of dense hydrogen.

Above analysis suggests NQE cannot results in continuous descent of melting temperature of $H_M$ within the pressure range from 500 GPa to 1.5 TPa. It also implies that some other thing might occur when above 1 TPa. In order to show this and to clarify the discrepancies with Ref.[19], as well as to further consolidate above conclusions, we turn to the direct two-phase simulations in the next subsection.

*2. Two-phase method estimation*

The computation of [19] suggested that the $T_m$ of $H_M$ beyond 1 TPa might be below 50 K, which is inconsistent with above analysis. Their calculation did not answer the question of whether $T_m$ approaches absolute zero or not, nor whether the destabilization of the solid phases is due to thermal noise or just because of NQE. On the other hand, our analysis presented above suggests strong stability of solid phases and the diminishing of driving forces at low temperatures. By contrast, in Ref. [19] the structural relaxation was reported to equilibrate very rapidly. This is inconsistent with the scenario suggested by Fig.4 too. Considering the two-phase method in *NVT* ensemble as used in Ref. [19] is prone to ambiguous results, especially at low temperatures, one might be suspicious about its conclusion [40]. In order to address these discrepancies, we repeat the two-phase AI-PIMD simulations in *NVT* ensemble using 32 beads, the same as in Ref.[19]. To reduce the impact of residual stress and energy on the results, three additional strategies are employed: (*i*) using a large cell with 480H/cell, rather than the 200H/cell as in Ref.[19]. This allows more flexible distortions to dissipate the stress and strain energy; (*ii*) relaxing the initial two-phase coexistent configurations using AI-PIMD with the mass-centers being fixed, so as to remove the residual stress and energy largely; (*iii*) at the initial stage of the full AI-PIMD simulations, a small time step of 0.2 fs was used to increase the integration accuracy of the equations of motion, which is effective in reducing the unwanted non-equilibrium disturbances to the system.

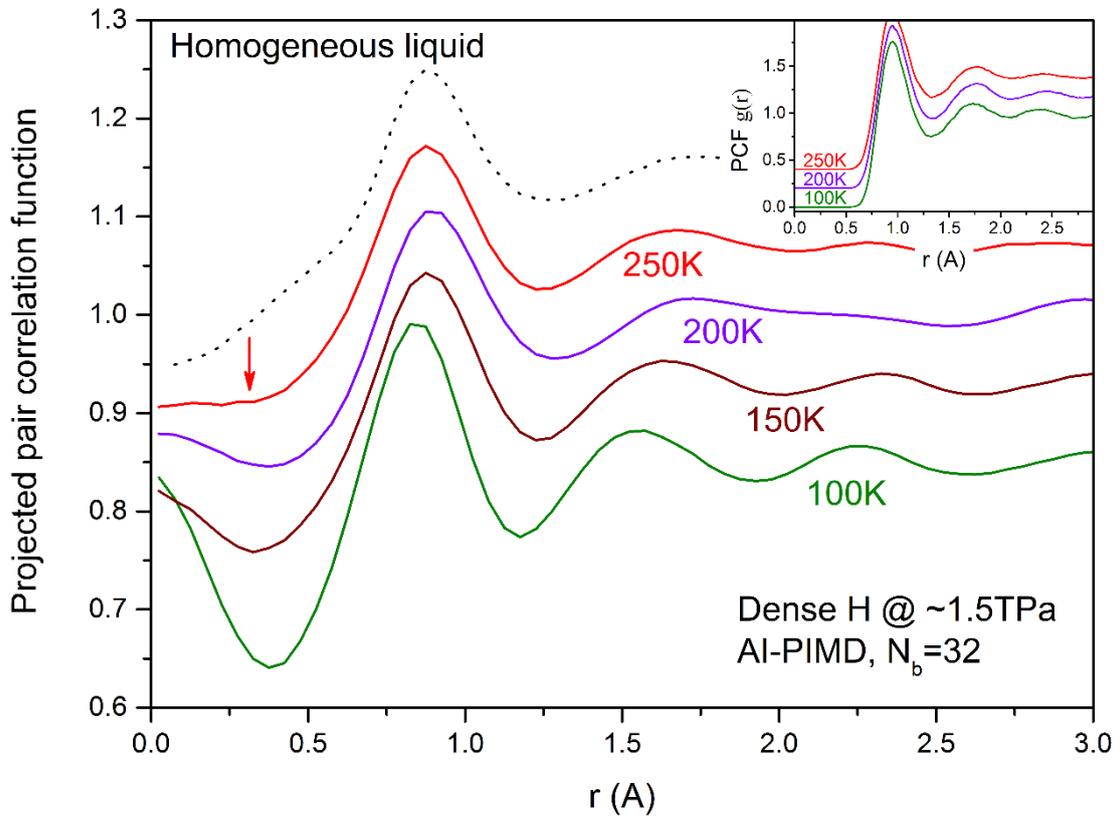

FIG. 5: (color online) Variation of the projected PCF in $H_M$ with temperature at a pressure

of ~1.5 TPa, after a long (~6 ps) two-phase equilibrating. The arrow indicates the beginning of the homogeneous liquid feature. Lines are relatively shifted for presentation.

The results of the two-phase simulations are surprising, though not totally unexpected. Different from Ref.[19], we find that $H_M$ liquefies smoothly and rapidly only at temperatures higher than 250 K. At lower temperatures, the two phases are found to coexist for a long time. Even after the initial solid phase already becomes unrecognizable, the system still requires a long time to equilibrate. This observation is totally in line with the picture implied in Fig.4: low temperature reduces the thermal driving forces, thus hinders the transformation among configurations, whereas the quantum motion of nuclei is not enough to destabilize the structure. For most of these simulations, the final equilibrated state is not the true homogeneous liquid. Rather it is an intermediate state between liquid and solid phase. On one hand, it is very similar to the liquid phase, both by visual identification and commonly used inspection tools, such as the angularly averaged PCF as shown in the inset of Fig.5 where the calculated $g(r)$ for three different temperatures are almost identical and show typical liquid features, as well as the mean square displacement (not shown). On the other hand, it has anisotropy and some long range ordering, being analogous to a solid. Figure 5 plots the projected PCF along *Z* direction, by comparison to that of the homogeneous liquid. The projected PCFs along *X* and *Y* directions are similar to the homogeneous liquid one, thus are not shown here. The drastic difference between these projected PCFs indicates that this phase is anisotropic when below 250 K. Also note that the peaks at the null projection distance reveal that in this phase the particles prefer to align along *Z* direction, a kind of long range ordering. Therefore we can confidently conclude that this phase is not a homogenous liquid, and $H_M$ does not melt at these thermodynamic conditions.

Figure 5 suggests that this exotic phase melts to the homogeneous liquid at about 250 K. As the temperature decreasing, its difference from that of the true liquid becomes more striking. A similar state was also observed in *heat until melting* simulations of the *Fddd* phase at ~1 TPa and 1.5 TPa (termed as phase *D* [40]). So why Ref.[19] obtained a low temperature liquid state down to 50 K in their two-phase simulations? The most plausible explanation is that they mistook this exotic phase as the true liquid, since they employed the traditional angularly averaged PCF and mean squared displacement that are incapable to distinguish these two phases. That is, failed to detect this fluid-like but non-liquid state and mistaking it as the true liquid might be the main reason that led Chen *et al.* to claim an ultra-low $T_m$. One additional strong evidence that supports this argument is that in our two-phase simulations this exotic phase is found stable down to below 100 K. On the other hand, our two-phase simulation also reveals that the solid *Fddd* phase is favored at 50 K [40], thus predicting a phase boundary between them at about 75 K under ~1.5 TPa. As Fig.3 showing, the DFT setting of Ref.[19] underestimated the stability of Cs-IV phase by 50~100 K. Hence it is very possible that in their calculations this phase boundary is pushed down below than 50 K. As for the fast equilibrating they observed in $H_M$, it can be readily explained by their uncontrolled two-phase simulation in *NVT* ensemble, where artificial driven forces accelerate the relaxation process.

**C. Stability of solid phase when approaching 0 K**

Our two-phase simulations suggested that the solid *Fddd* phase is favored against the liquid at low temperature, and it transforms into the fluid-like but anisotropic state with long-range ordering when above 75 K, the latter then melts to the homogeneous liquid state at about 250 K when at a pressure of ~1.5 TPa. This is consistent with the estimation of $T_m$ given in Fig.4. We will provide another strong evidence that further supports this picture below. That is, we are going to evaluate the relative stability of the solid and liquid phases when approaching 0 K. To address this issue, usually one will resort to the principle of minimal free energy to determine which one is the favored phase. Calculation of free energy is cumbersome and computation demanding. We thus take the advantage of the fact that the internal energy of both the harmonic and anharmonic phonons of $H_M$ already converge to their respective zero point energy, and the difference between zero point energy and the free energy is less than $10^{-5}$eV/H at 50 K [40]. Please note that even though the harmonic approximation is very crude for dense hydrogen, the qualitative magnitude of its internal energy and free energy are nevertheless reliable. For this reason, we can simply compare the enthalpy at 50 K to assess the relative stability of the liquid and solid phases as zero Kelvin is approached.

The enthalpies calculated with AI-PIMD using 32 and 64 beads, and the extrapolation to infinite number of beads [41], are shown in Fig.6 for both the liquid and solid *Fddd* phases, respectively. It may be seen that the solid phase is always favored over the liquid one. The enthalpy difference is about 16 *m*eV/H in the case with 32 beads, and decreases to 14.7 *m*eV/H when using 64 beads. The converged result obtained by Richardson extrapolation [41] is 14.3 *m*eV/H. Therefore we find that inclusion of the quantum motion of protons does not confer the liquid phase much advantage. This observation establishes the first direct numerical evidence at the DFT level that dense hydrogen is actually in a solid groundstate when at around 1.5 TPa.

From our extensive AI-PIMD calculations for the solid phases at 50 K, it seems unlikely that there will take place a spontaneous delocalization of the nuclear wavefuntion at lower temperatures. In the cases we have studied, the dispersion of the integral paths in solid phases is confined mainly by the potential well, rather than by the de Broglie thermal wavelength. Therefore lowering the temperature might not enhance the tunneling probability very much. Inclusion of the identical particle exchanges is also unlikely to change the melting temperature qualitatively, since the contribution of exchanges to the free energy of the liquid phase is expected to be small (*e.g.*, it is at a level of ~1 K for $^4$He [25]), whereas the enthalpy difference between the solid and liquid phases of $H_M$ is greater than 160 K.

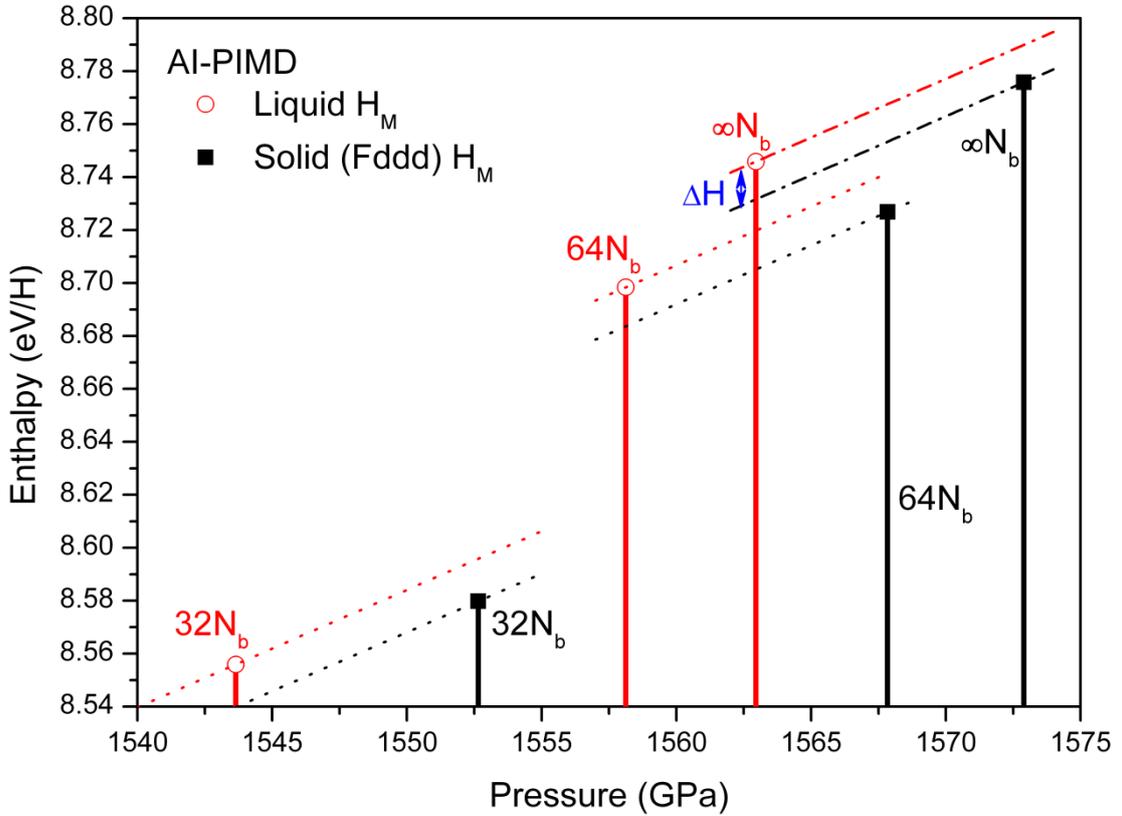

FIG. 6: (color online) Comparison of the enthalpy of *Fddd* and the liquid phase of $H_M$, calculated using AI-PIMD at 50 K with 32 and 64 beads, respectively, and the extrapolated results to the infinite beads. Dotted and dash-dotted lines extrapolate the enthalpy to nearby pressures.

## IV. CONCULSION

In summary, by decomposing the descent of the melting temperature of $H_M$ into two separate issues, *i.e.*, the interaction potential softening and the dynamic driving forces from thermal noise or quantum ZPM, we presented a complete solution for the former by using AIMD simulations and the Z-curve method to evaluate the classical superheating limit and the melting curve of $H_M$. The second issue was also addressed by using AI-PIMD simulations, which revealed that inclusion of NQE would lower the superheating limit and melting curve accordingly. Within the pressure range from 500 GPa to 1.5 TPa, the groundstate of dense hydrogen was predicted to be solid rather than the conjectured liquid. The melting/destabilizing mechanism of the crystalline phases was determined to be thermal activation. The melting temperature was estimated to be 200~250 K and has a flat variation with pressure. This provides a completely distinct picture about $H_M$, and defies the continuous descent of the melting temperature that was claimed previously.

**Acknowledgement**

This work was supported by the National Natural Science Foundation of China under Grant No.11274281, the CAEP Research Project 2012A0101001, the fund of National Key Laboratory of Shock Wave and Detonation Physics of China (under Grant No. 9140C670105130C67237), the National Science Foundation through Grant No. CHE-0910623, and also by EFree, an Energy Frontier Research Center funded by the U.S. Department of Energy (Award No. DESC0001057 at Cornell). Computation was performed using the Extreme Science and Engineering Discovery Environment (XSEDE), which is supported by National Science Foundation grant number OCI-1053575, the Cornell NanoScale Facility, a member of the National Nanotechnology Infrastructure Network, which is supported by the National Science Foundation (Grant ECCS-0335765), and the resources of the Supercomputing Laboratory at King Abdullah University of Science & Technology (KAUST) in Thuwal, Saudi Arabia.

# Lattice stability and high pressure melting mechanism of dense hydrogen up to 1.5 TPa

Hua Y. Geng,[1,2] R. Hoffmann,[2] Q. Wu[1]

[1] National Key Laboratory of Shock Wave and Detonation Physics, Institute of Fluid Physics, CAEP; P.O.Box 919-102 Mianyang, Sichuan, P. R. China, 621900

[2] Department of Chemistry and Chemical Biology, Cornell University, Baker Laboratory, Ithaca, New York 14853, USA

# Supplementary Information

## A. Convergence of k-points

Figures S1 and S2 show the convergence of the energy and pressure with respect to different k-point meshes. The configurations were sampled from AIMD simulations for both liquid and solid phases. Note that the convergence is achieved with a k-point mesh of $3 \times 3 \times 3$.

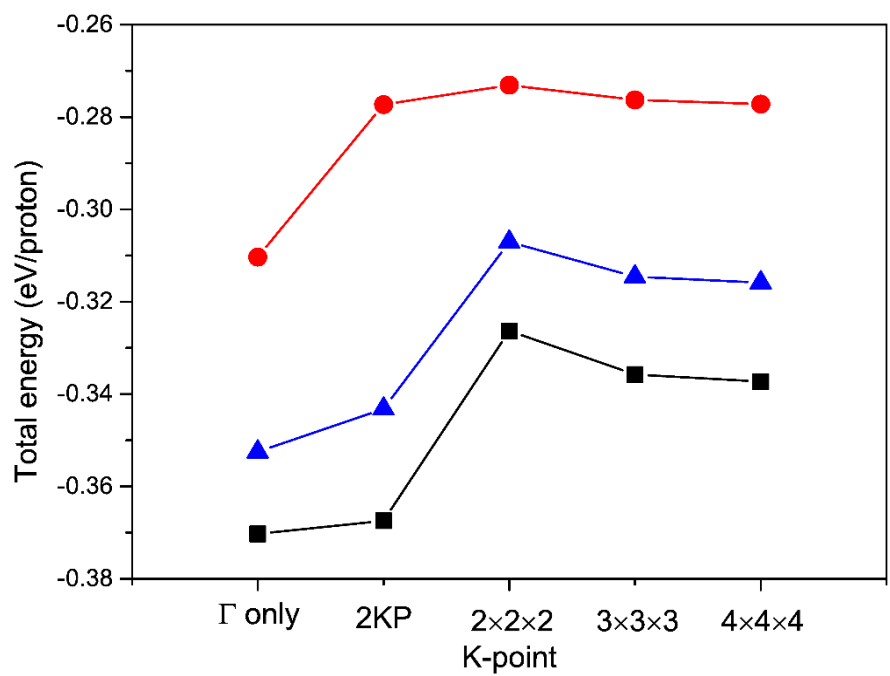

FIG. S1: Convergence of the DFT total energy with respect to k-point mesh size. The configurations were sampled from AIMD simulations for both liquid and solid phases.

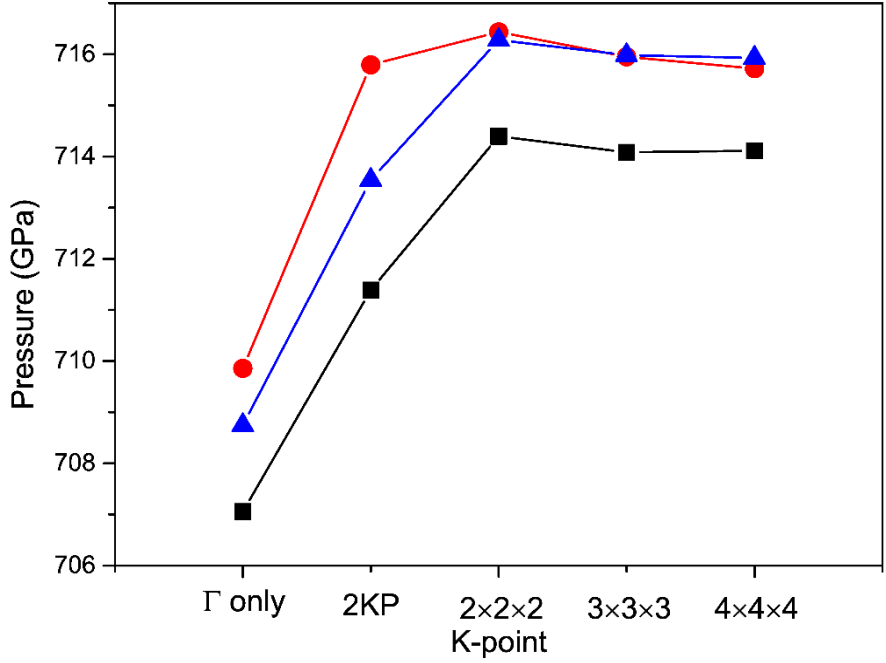

FIG. S2: Convergence of the pressure with respect to k-point mesh size. The configurations were sampled from AIMD simulations for both liquid and solid phases.

**B. Phonon density of state**

The phonon spectra in a harmonic approximation were calculated with the small-displacement method, as implemented in the PHON software [1]. Sufficiently large supercells, containing 128 atoms, were used. In the associated DFT calculations, a Brillouin zone sampling mesh of $11 \times 11 \times 11$, a kinetic energy cutoff of 1000 eV, and a very dense support augmentation charge grid that is required for an accurate force calculation were used. This setup gives a convergence in the ZPE better than 2 *m*eV/H. The magnitude of the small displacement was slightly varied to check the numerical stability of the calculated force constant matrices. The ZPE was estimated from the phonon density of states $g(\omega)$ by $\int g(\omega)\, \hbar\omega/2\, d\omega$.

The anharmonic effect in the phonon density of states was roughly estimated using AIMD simulations: at first an NVT MD simulation was run to equilibrate the system at 50K, which is then followed by a long NVE simulation lasting for 5 ps. The velocity auto-correlation function was then extracted from the MD trajectory. Fourier transformation of this function then gives the phonon DOS that contains anharmonic contributions [2-4]. The obtained DOS at 700 GPa for *Fddd* phase is shown in Fig. S3. It can be seen that anhamonicity enhances the low-frequecy modes and reduces those at high-frequency region. Both harmonic and anharmonic phonon DOS were used to estimate the thermal contribution to the internal energy and free energy. Though anharmonicity changes the ZPE by about 0.1 eV/H, both harmonic and anharmonic phonon DOS predict that at 50 K the free energy and ZPE are almost equal, with a deviation less than $10^{-5}$ eV/H. In liquid phase the local vibrations have similar magnitude, thus the same conclusion also holds. This observation allows us to assess the relative stability of the liquid and solid phases when approaching 0 K by using the enthalpy at 50 K, which is accessible easily in PIMD simulations.

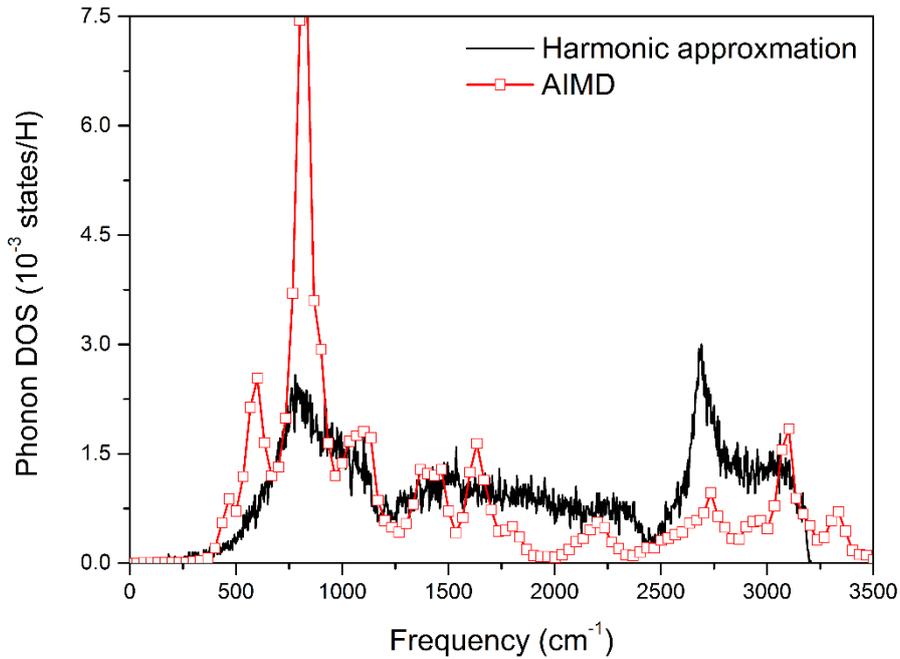

FIG. S3: Comparison of the phonon DOS of *Fddd* phase at 700 GPa calculated with harmonic approximation and AIMD simulations, respectively.

## C. Structural relationship between Cs-IV and *Fddd* phases

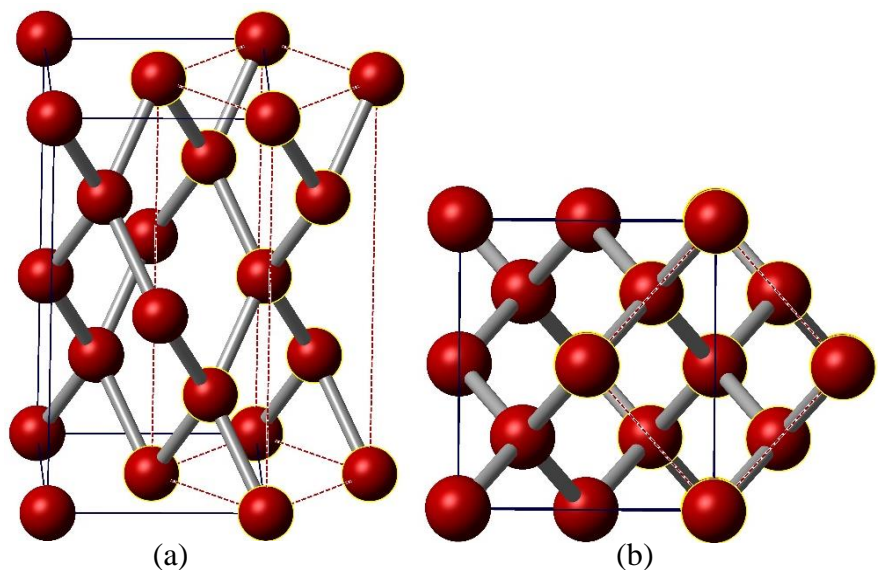

(a)  (b)

FIG. S4: Structural connection between the Cs-IV and *Fddd* phases: (a) side view and (b) top view. The solid dark lines denote the unit cell of the *Fddd*, and dashed red lines are those of the (distorted) Cs-IV, both are low symmetry distortions of the cubic diamond structure.

Figure S4 illustrates the structural relationship between the Cs-IV and *Fddd* phase [5]. Each unit cell of *Fddd* contains two units of Cs-IV. Both structures are derived from the cubic diamond phase: a tetragonal deformation of the cubic diamond cell (by increasing *c/a* ratio) leads to Cs-IV, from which the *Fddd* will be obtained if one also increases the *b/a* ratio (*i.e.*, an orthorhombic distortion). This distortion alters the nearest neighbor distance of protons, as shown in the histogram of Fig.S5, from which it is clear that while the two structures are related to each other by a distortion, the atomic local environment are not the same.

It can be seen that in *Fddd* phase hydrogen atoms have a slightly shorter nearest neighbor separation, below $1.0\,\text{Å}$. Thus the *Fddd* phase might have a stronger "residual" chemical interaction, a memory of the molecular form of H as $H_2$. In AIMD simulations, we observed a transition from Cs-IV to *Fddd* when 2KP (two special k-points) was used. But no inverse transition has been observed. The appearance of the Cs-IV→*Fddd* transition indicates that the absence of the structural oscillation in this system is not a consequence of the finite size of the simulation cell.

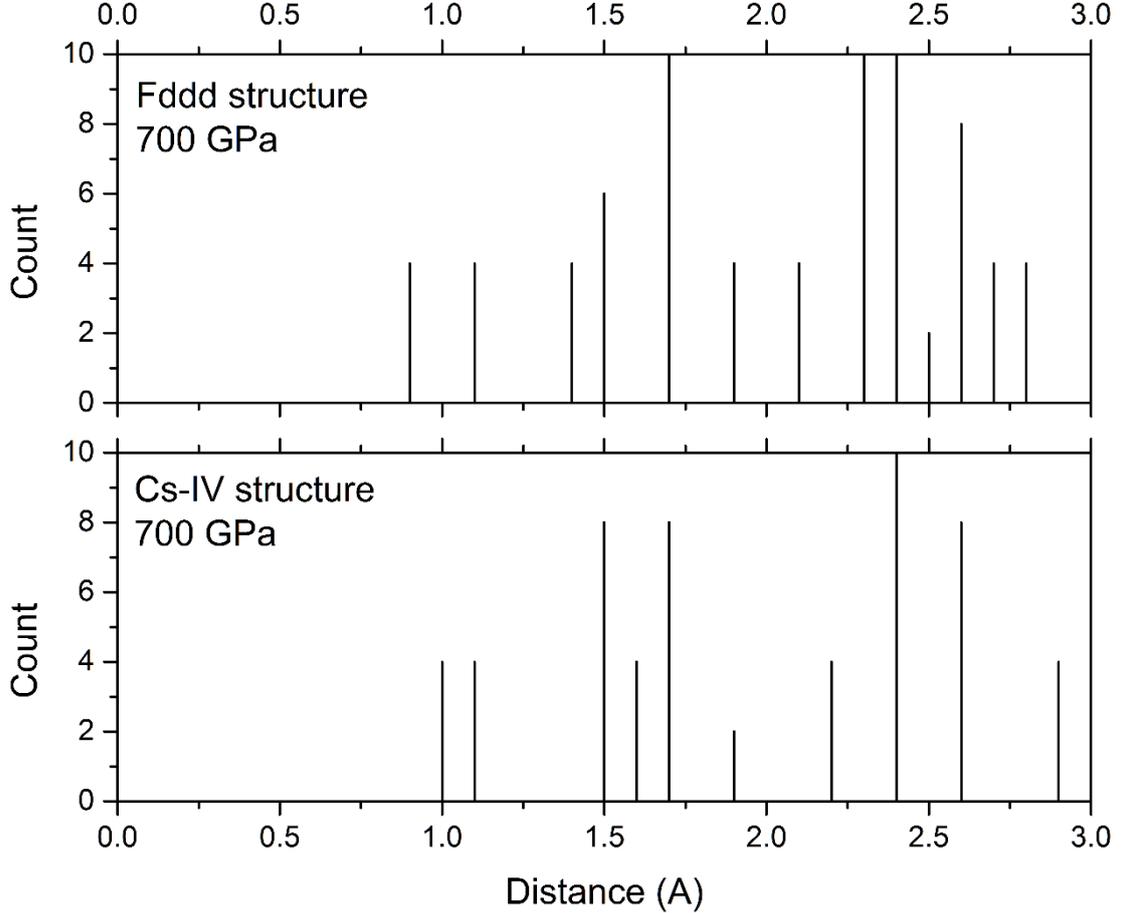

FIG. S5: Internuclear separation histogram of *Fddd* and Cs-IV phases of hydrogen at ~700 GPa, respectively.

**D. Structure and ordering of the phase *D***

[Note: all calculations of the phase *D* as discussed below were performed with the 2KP for the k-points sampling.]

Usually in numerical simulation we detect the melting or solid-solid transition by inspecting the variation of the angularly averaged pair correlation functions (PCF) of atoms. With this function we can get some insight about the atomic environment, and thus the physical state of the system. Taking the melting of *Fddd* phase at about 740 GPa calculated using AI-PIMD with 2KP as an example, the calculated PCFs of proton mass centers are shown in Fig.S6. It can be seen that the long-range correlations between protons persist up to 450 K, and disappear at 500 K. Therefore we can conclude that the melting temperature is between 450 and 500 K (Note that with a converged k-points mesh, the melting temperature will drop to below ~300 K). Though this method works well for most materials, it is not generally valid and can fail in some special cases. The phase *D* of dense hydrogen is one of them.

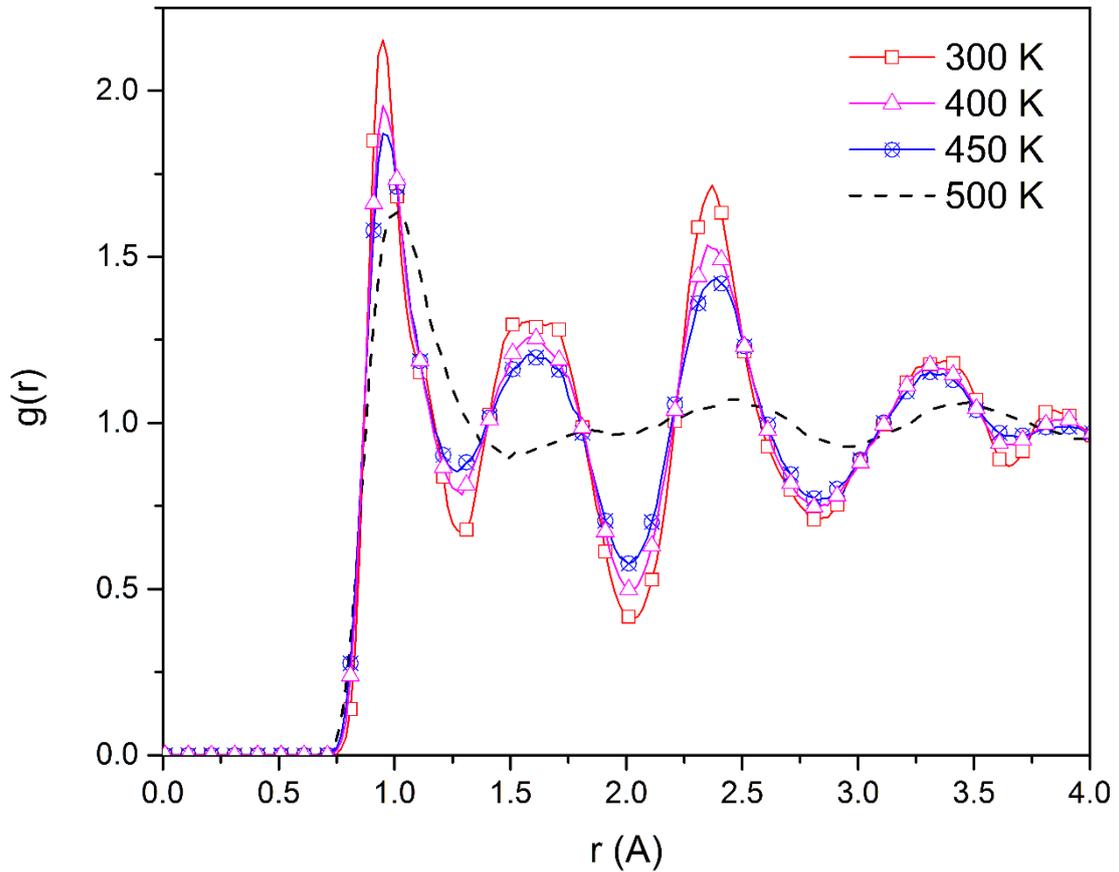

FIG. S6: Angularly averaged PCF of the proton mass centers of dense hydrogen at about 740 GPa simulated using AI-PIMD with 2KP. The results are obtained by averaging over a time scale of 1.0 ps, after a structural equilibration for about 2.0 ps.

Unlike a traditional solid, phase *D* does not have a well-defined static lattice. Its long-range ordering is revealed from long-time motion statistics. Therefore, the static appearance or symmetry of a snapshot of the atoms in this phase might be quite different from the real positional ordering, and might change drastically from time to time. This exotic phase was found in AI-PIMD simulations with 2KP at around the classical melting line of the *Fddd* phase of hydrogen. The importance of this phase is that it undermines the stability of Cs-IV and *Fddd* greatly when above 1 TPa, thus might lead to a mistake in determination of the melting temperature.

In this new phase, the particles (hydrogen atoms) are mobile and diffuse. This feature is quite similar to a liquid, and both short time motion trajectories and static snapshots of particle positions show that the phase could well be mistaken for a liquid. For example, the widely used indicator for melting, the angularly averaged PCF *g*(*r*), has an appearance similar to that of a liquid for the phase *D* we just discovered, as can be seen in the inset of Fig.S7 for dense H at 500 K and around 1.1 TPa. Figure S8 shows two consecutive snapshots separated by 50 fs in time, taken from an AI-PIMD simulation. The snapshots do not show common local features, and no correlation or ordering can be easily observed. Nevertheless, after calculating long-time statistics (over about 3 ps), the distribution of proton density incredibly unfolds a long-range ordering (the last panel in Fig.S8).

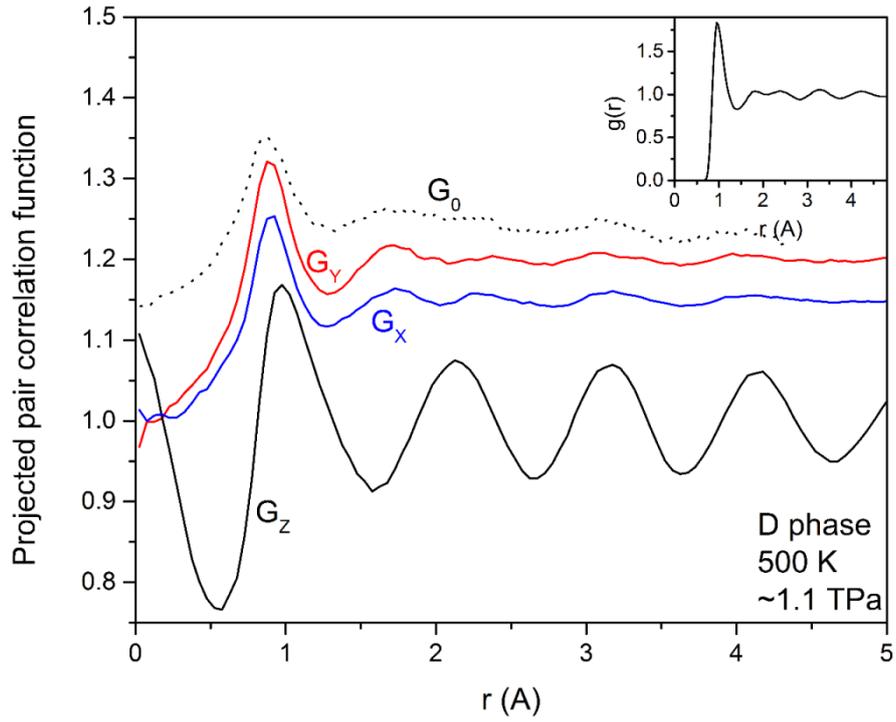

FIG. S7: Azimuthal angle averaged PCF $G(r)$, projected along Cartesian $X$, $Y$, and $Z$ direction of the simulation cell, respectively, for the phase $D$ of dense H at 500 K and ~1.1 TPa. Inset: 3D angularly averaged PCF $g(r)$, for which a projected PCF that assumed a homogeneous liquid is also plotted and denoted as $G_0$ for the purpose of comparison.

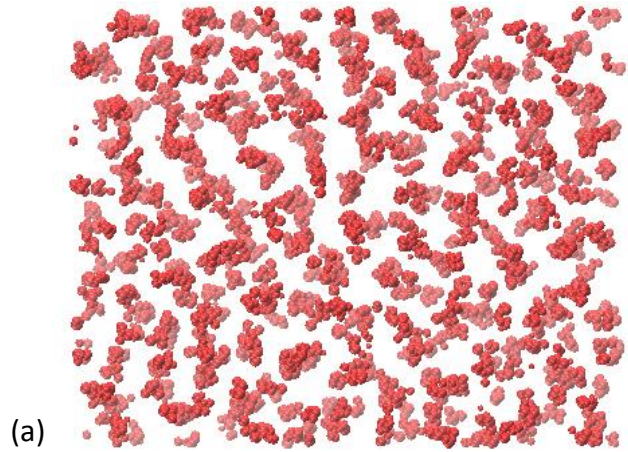

(a)

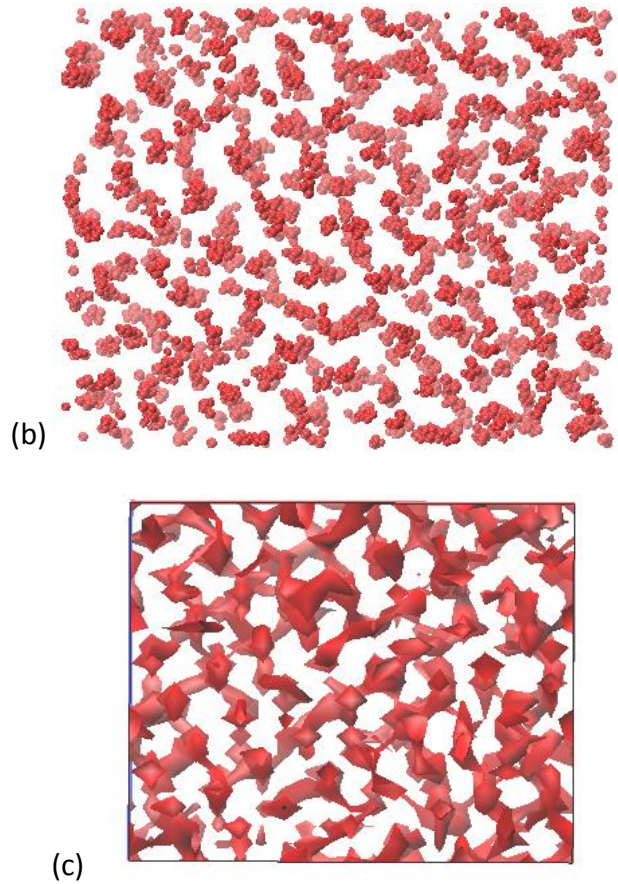

FIG. S8: Top view (along Z direction) of two consecutive snapshots separated by 50 fs [(a) and (b)], and their long-time (3 ps) averaged proton density isosurface [(c)] taken from an AI-PIMD simulation after 3 ps equilibration of phase D at 500 K and ~1.1 TPa. Notice the emergence of long-range positional ordering in the last panel.

When viewed from other two perpendicular directions (side views), as shown in Fig.S9, the ordering becomes less evident. This difference reveals the anisotropy of the system; the details, of course, depend on the orientation of the simulation cell. By rotating the side view angle of about 45 degrees, we can find a clearly layered ordering. But no other ordering can be found; therefore the dynamic long-range ordering in this phase can be termed as two-dimensional.

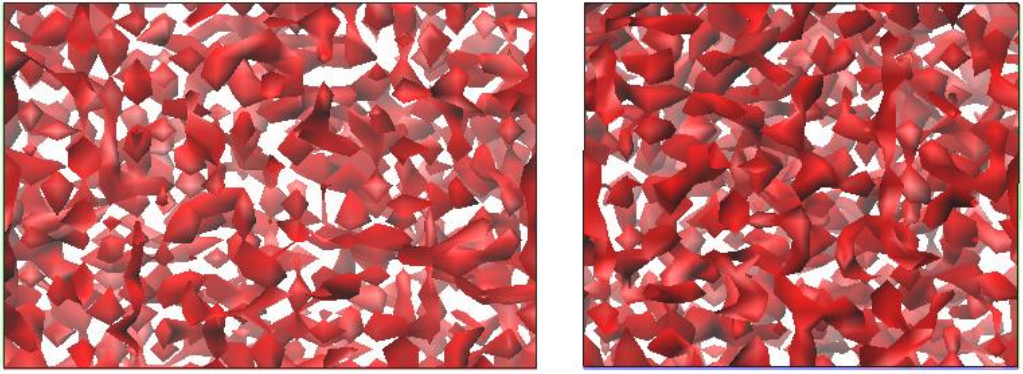

FIG. S9: Side views (along *X* and *Y* directions) of the isosurface of the proton density that was averaged over 3 ps of an AI-PIMD run for the phase *D* at 500 K and ~1.1 TPa.

Since the usual isotropic PCF fails to distinguish this dynamic phase *D* of hydrogen from a liquid, we need a new mathematical tool to analyze the partial ordering in this structure. Inspired by the anisotropy of the proton density, as shown in Figs.S8 and S9, we introduce the projected PCF *G*(*r*): First project all particles onto a plane, and then carry out a statistical analysis to evaluate their two-dimensional PCF in that plane (for details see the main text). Such projected PCFs have the capability to unveil the underlying anisotropy, as well as to differentiate phase *D* from a liquid.

The power of *G*(*r*) is illustrated in Fig.S7, where the projections have been done along three perpendicular Cartesian directions. $G_Z(r)$, the projected PCF along the *Z* direction, shows clearly a strong long-range ordering even at 500 K, which is about 100 K higher than the classical $T_m$ of the *Fddd* phase at 1.1 TPa when calculated using two special k-points. Note that there is a striking peak at the null projected distance in $G_Z(r)$, which results from atoms aligning along the Z direction—a strong indication of positional ordering. On the other hand, projection along directions perpendicular to *Z*, $G_X(r)$ and $G_Y(r)$, show few features after the nearest neighbor peak.

The underlying anisotropy of phase *D* is thus clearly revealed. More importantly,

if the system is a homogeneous liquid, there is a simple relation between $g(r)$ and $G(r)$, which can help us to identify whether a system already melted or not. In Fig.S7, the $G_0(r)$ is calculated from the $g(r)$ by assuming the system as a homogeneous liquid, and using the relation given in the main text. Its key feature is the monotonic increase of $G(r)$ to the first peak, then quickly growing featureless at larger distances. Obviously, $G_X(r)$ and $G_Y(r)$ look more like a liquid, but $G_Z(r)$ definitely not. Therefore, we conclude that under this thermodynamic condition, phase $D$ is not molten.

**E. Ambiguity and sensitivity of two-phase method in *NVT* ensemble**

The two-phase method is widely employed in MD or Monte Carlo simulations of first-order phase transition, especially when modeling melting [6-8]. The main purpose of this approach is to eliminate the hysteresis-related superheating or supercooling phenomenon near the phase boundary, which can eventually be traced to the energy barrier that separates the corresponding phases. The idea behind the method is that by artificially providing condensation nuclei to the otherwise homogeneous system, the global symmetry is broken and a nucleation-and-growth mechanism of phase transition lowers the energy barrier greatly. Practice has showed that this method is effective in removing superheating/supercooling effects [6-8].

However, when this method is used in *NVT* ensemble, we find that the results might be ambiguous. The reason is that by creating condensation centers or a coexistence interface, local stress is unavoidably introduced into the system. Taking the melting case for example, the residual stress will drive the system towards the perfect solid phase (or the liquid, depending on how the stress is distributed over the simulation cell) to reduce the strain energy. The consequence is that the simulated melting temperature might be changed, sometimes up to several hundred Kelvin.

Another factor that will affect $T_m$ is the residual internal energy of the liquid part in the two-phase configurations. High residual energy (sampled from a high temperature liquid) will make the system molten before equilibrium, and a too low residual energy (sampled from a low-temperature liquid) will freeze the system prematurely, thus invalidating the assumption of the two-phase method. Though, in principle, one can tune the thermostat to control how fast the excess energy is being removed from the system, we find that this process is tricky when one works in an *NVT* ensemble.

Fortunately all of these problems are naturally solved in the *NPT* ensemble, in which the fluctuations of the cell volume and shape dissipate effectively the residual stress and excess internal energy of the liquid part, and the results become almost insensitive to the initial state. This is not the case for the *NVT* ensemble: the residual stress due to the solid/liquid interface and the excess energy in the liquid part affect the final results strikingly.

Figure S10 shows the ambiguity in the results of the two-phase method in *NVT* ensemble when applied to a supercell with 200 hydrogen atoms in it (the same setting as [9]). It can be seen that the estimated melting temperature (using AIMD) is sensitive to the initial conditions: initially half of the atoms of point *A* (the liquid part is sampled from a liquid equilibrated at 3000 K) and *B* (the liquid part is sampled from a liquid equilibrated at 900 K) are liquid, whereas the solid/liquid ratio is 3:5 for point *C* (for which the liquid part is sampled from a 900 K liquid). Point *D* is taken from [9] and the condition of the liquid part is unknown. These data are scattered, which clearly illustrates that the $T_m$ estimated using *NVT* and the two-phase method can be anywhere between the superheating and supercooling limit. There are some extreme cases, like point *C*, where the system freezes directly into a quasi-glass state in which atoms still have some mobility and thus is difficult to distinguish it from the true liquid.

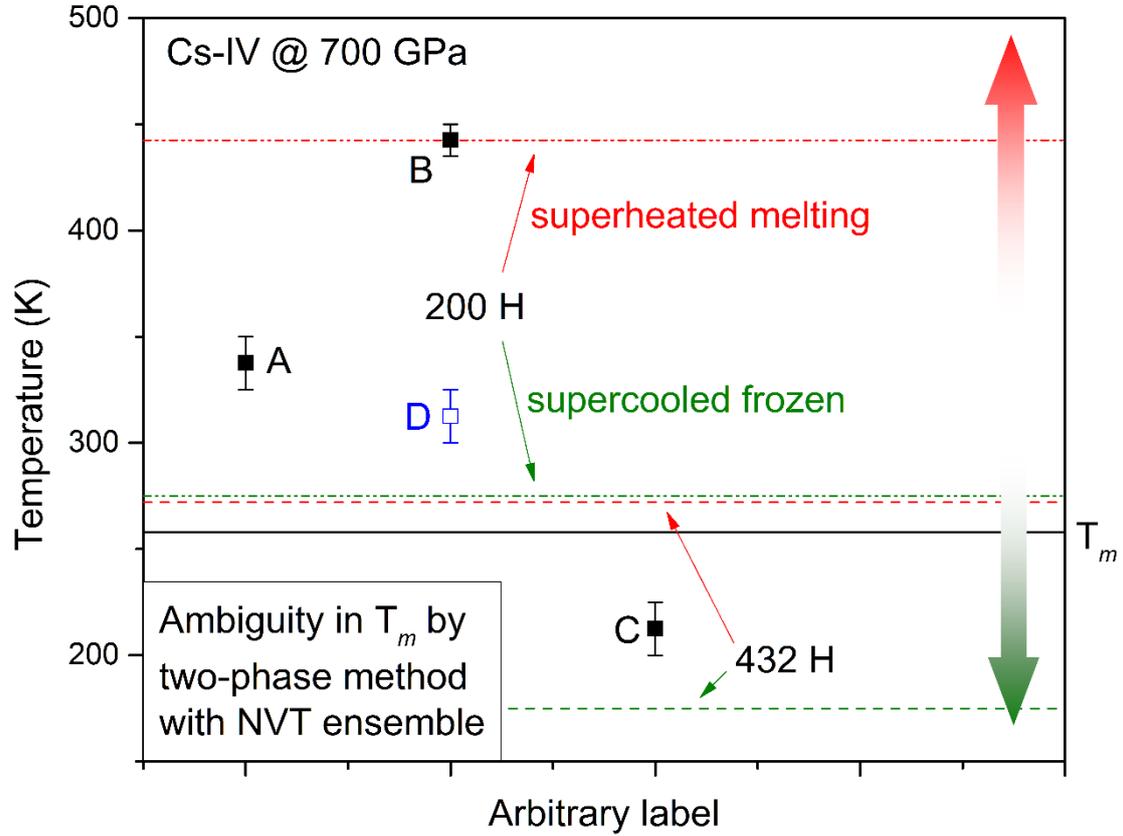

FIG. S10: Ambiguity in classical melting temperature calculated using the two-phase method in *NVT* ensemble: points *A*, *B*, *C* are by this study, and point *D* is from [9]; all are for a supercell having 200 H in it. Also shown are the superheating and supercooling boundaries for a cell with 200 H and 432 H, respectively. The solid line denotes the $T_m$ of a cell with 432 H in it that is calculated using the Z-curve method and with 2KP.

Besides the equilibrated temperature of the liquid part and the solid/liquid ratio, the results also depend on the crystal orientation of the solid part and the geometry of the solid/liquid interface. Since all calculations performed in [9] employed the two-phase method in *NVT* ensemble, we suppose that they might also suffer from these shortcomings of the methodology.

In order to alleviate the above mentioned pathology in *NVT* two-phase method, we prepare the initial two-phase coexistent configurations by separately equilibrating the solid and liquid part to the target temperatures. In addition, three strategies in AI-PIMD simulations are employed: (*i*) using a large cell with 480H/cell, rather than the 200H/cell as in [9]. This allows more flexible distortions to dissipate stress and strain energy; (*ii*) relaxing the two-phase coexistent initial configurations using AI-PIMD with the mass-centers being fixed, so that to remove the residual stress and energy largely; (*iii*) at the initial stage of the full AI-PIMD simulations, a small time step of 0.2 fs was used to increase the integration accuracy of the motion equations, which is effective in reducing the unwanted non-equilibrium disturbances to a minimal level. This treatment improves the reliability of the *NVT* two-phase simulations. There are two fundamental findings beyond the calculations in [9]. At first, the melting (or destabilization) of the solid phase of $H_M$ is found to be very slow, instead of the rapid melting (1~2ps) as claimed in [9]. It takes about 3ps to reach the equilibrium state when at 200 K and ~1.5 TPa, and becomes much slower with decreasing temperatures. Especially, the two-phase coexistence lasts for more than 4.5ps when at 50 K, and then a sign of relaxing towards the solid phase appears, as shown in Fig.S11. It should be noted that the two-phase simulation at 50 K finally restores the *Fddd* phase at about 10 ps. The slow destabilizing process implies that the melting of $H_M$ is mainly driven by thermal activation, rather than by nuclei quantum effects, such as tunneling. The fast melting observed in [9] can be explained by the residual energy or stress, which provides artificial driving forces to the system.

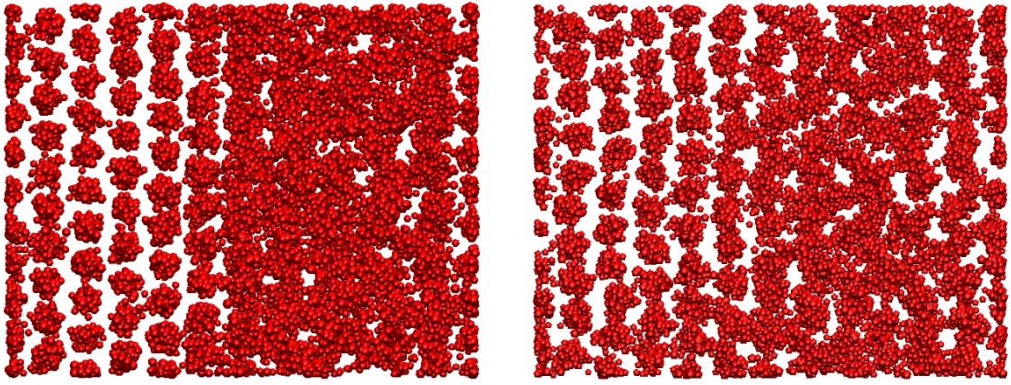

FIG. S11: Snapshots taken from two-phase AI-PIMD simulations with 32 beads at 50 K and ~1.5 TPa. Left: the initial configuration; right: after 4.5 ps equilibrating. Note there is a sign towards solidifying.

Another finding is that $H_M$ does not melt directly into a true liquid. Rather, it transforms from *Fddd* to a phase *D*-like state at 75 K and ~1.5 TPa. We cannot confirm that this state is really phase *D* or not, but their behavior is very similar, as revealed by the projected PCFs shown in Fig.5 of the main text. Alternatively, it could be a glassy state, with some mobility. Figure S12 shows a snapshot of an equilibrated configuration taken from the two-phase AI-PIMD simulations of 150 K at about 6 ps. Some non-liquid structural features still can be observed, corroborating that $H_M$ in this state is not molten. These observations support the argument that $H_M$ does not melt at these conditions. The "liquid state" beyond 1 TPa obtained in [9] might be a phase *D* like state, instead of being the true liquid.

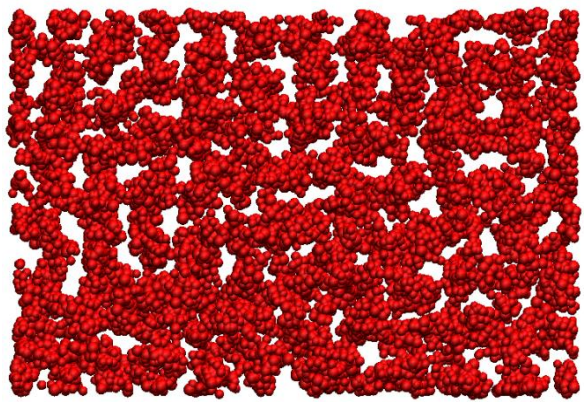

FIG. S12: Snapshot taken from equilibrated two-phase AI-PIMD simulations with 32 beads at 150 K and ~1.5 TPa, at the simulation time of 6 ps. Some non-liquid structural features still can be observed.

**Supplementary references**